\documentclass[preprint,onecolumn,aps,floats]{revtex4}

\usepackage{graphicx}

\begin{document}

\title{Experimental Realization of Entanglement Concentration and A Quantum Repeater}

\author{Zhi Zhao$^1$, Tao Yang$^1$, Yu-Ao Chen$^1$, An-Ning Zhang$^1$ and Jian-Wei Pan$^{1,2*}$}
\address
{$^1$Department of Modern Physics, University of Science and
Technology of China, Hefei, Anhui 230027, China\\
$^2$ Institut f\"{u}r Experimentalphysik, Universitat Wien,
Boltzmanngasse 5, 1090 Wien, Austria}
\date{December 21, 2002}

\begin{abstract}

\ We report an experimental realization of entanglement
concentration using two polarization-entangled photon pairs
produced by pulsed parametric down-conversion. In the meantime,
our setup also provides a proof-in-principle demonstration of a
quantum repeater. The quality of our procedure is verified by
observing a violation of Bell's inequality by more than 5 standard
deviations. The high experimental accuracy achieved in the
experiment implies that the requirement of tolerable error rate
in multi-stage realization of quantum repeaters can be fulfilled,
hence providing a practical toolbox for quantum communication
over large distances.

\
\

\end{abstract}

\maketitle

In recent years, significant experimental advances have been
achieved in the field of quantum communication (QC). In
particular, entangled photon pairs have been used to
experimentally demonstrate a number of QC schemes such as dense
coding \cite{bennett92,mattle96}, quantum teleportation
\cite{bennett93,dik97,martini98,pan03} and quantum cryptography
\cite{bb84,ekert91,thomas00,gisin00,kwiat00}. Though the above
schemes are achievable for moderate distances (up to a few tens
of kilometers in quantum cryptography), serious problems occur in
bringing QC to technologically useful scales. One of the problems
is the photon losses in the transmission channel. In quantum
cryptography, the photon losses by themselves only reduce the bit
rate (exponentially with distance). With perfect detectors the
distance would not be limited, one can solve this problem by
sending more photons. However, due to the dark counts, whenever a
photon is lost there is a chance that a dark count produces an
error. Hence, when the probability of a dark count becomes
comparable to the probability that a photon is correctly
detected, the signal-to-noise ratio tends to 0. Fortunately, as
shown in ref. \cite{gisin02} the use of entangled photons and of
entanglement swapping \cite{marek93} could offer ways to extend
the achievable distances.

Another problem is caused by the noisy quantum channel. The
decoherence degrades the quality of entanglement between two
particles more and more the further they propagate. However, all
the above protocols require that two distant parties, usually
called Alice and Bob, share highly entangled photon pairs. The
decoherence can be overcome by exploiting entanglement
purification \cite{bennett96a,deutsch96} - a way to extract a
subset of highly entangled states from a large set of less
entangled states using local operations and classical
communication. Therefore, quantum repeaters \cite{briegel98}, a
combination of entanglement swapping and entanglement
purification, hold the promise to solve the problems of the photon
losses and of the decoherence in long-distance QC.

Though entanglement swapping has been experimentally demonstrated
with high accuracy \cite{pan98,pan01a,thomas02}, the
implementation of quantum repeaters remains to be an experimental
challenge due to the difficulty to achieve entanglement
concentration. Various purification schemes have been proposed
for pure or mixed entangled states. They mainly include three
classes. First, the so-called Procrustean method
\cite{bennett96b}. It requires that the photon pairs are all in a
pure non-maximally entangled state, say, $\alpha\vert
H\rangle\vert V\rangle+\beta\vert V\rangle\vert H\rangle$. Here
$\alpha$ and $\beta$ are two {\em known} coefficients that satisfy
$\vert\alpha\vert^2+\vert\beta\vert^2=1$, and $\vert H\rangle$
and $\vert V\rangle$ denote horizontal and vertical polarizations
of the photons. In this case, the scheme only involves local
filtering operations on single pairs. Second, the so-called
Schmidt decomposition scheme \cite{bennett96b}, in which the
photon pairs are all in a pure but {\em unknown} non-maximally
entangled state $\alpha\vert H\rangle\vert V\rangle+\beta\vert
V\rangle\vert H\rangle$. In practice, this scheme is difficult to
implement, since it requires simultaneous collective measurements
on many photons. To distinguish the second case from the others,
in what follows we refer to it as entanglement concentration.
Third, the general purification method that works for arbitrary
mixed states \cite{bennett96a,deutsch96}. The implementation of
the general scheme is more difficult as it requires
controlled-NOT (CNOT) operations between different photons.

We emphasize that, though only the general scheme can be used to
purify arbitrary mixed states, both local filtering and
entanglement concentration are of interest in their own rights.
This is because, on the one hand, both methods provide a way to
generate maximally entangled states, which is different from the
general scheme where only highly entangled states are generated.
On the other hand, with the help of local filtering any
inseparable states can be purified \cite{horo97}, while the
general scheme alone only works for the cases where the
entanglement fidelity $F$ is larger than 1/2 \cite{bennett96a}.

Up to now, except for the Procrustean method has been demonstrated
very recently \cite{kwiat01}, there is no experimental
implementation of the other schemes due to the difficulty to
control many photons. To avoid the difficulties caused by
collective measurements and CNOT operations, several practical
schemes have been proposed for entanglement concentration
\cite{zhao01,imoto01} and entanglement purification \cite{pan01b},
where only linear optical elements are required. In the present
experiment, following the scheme as suggested in
ref.\cite{zhao01,imoto01}, we report the first experimental
demonstration of entanglement concentration. Meanwhile, using
local filtering \cite{kwiat01} and entanglement swapping
\cite{pan98} we are also able to report the first experimental
verification of a quantum repeater.

We first explain the entanglement concentration scheme (shown in
Fig. 1a). Suppose that two parties, Alice and Bob, would like to
share photon pairs in the maximally entangled state
$\vert\Psi^{+}\rangle=1/\sqrt{2}(\vert H\rangle\vert
V\rangle+\vert V\rangle\vert H\rangle)$. Further suppose that,
due to the imperfect quantum channel, what they share are two
photon pairs $\left( 1,2\right)$ and $\left( 3,4\right)$ in the
following unknown states:

\begin{equation}
\begin{array}{l}
\left\vert \Psi \right\rangle _{12}=\alpha \left\vert
H_{1}\right\rangle \left\vert V_{2}\right\rangle +\beta
\left\vert V_{1}\right\rangle \left\vert H_{2}\right\rangle,\\
\left\vert \Psi \right\rangle _{34}=\alpha \left\vert
H_{3}\right\rangle \left\vert V_{4}\right\rangle +\beta
\left\vert V_{3}\right\rangle \left\vert H_{4}\right\rangle,
\end{array}
\end{equation}
where Alice holds photons 1 and 3, and Bob holds photons 2 and 4.
Before concentration, the polarization of photon 4 is rotated by
$90^{\circ}$ using one half-wave plate (R$_{90}$ in Fig. 1a).
After passing through the half-wave plate, the state of photons 3
and 4 becomes
\begin{equation}
\left\vert \Psi \right\rangle _{34}=\alpha \left\vert H_{3}\right\rangle
\left\vert H_{4}\right\rangle +\beta \left\vert V_{3}\right\rangle
\left\vert V_{4}\right\rangle .
\end{equation}

Then Bob superposes photons 2 and 4 onto a polarization beam
splitter (PBS in Fig. 1a). Since the PBS transmits horizontal and
reflects vertical polarization, a coincidence detection between
the two outputs $2^{\prime}$ and $4^{\prime}$ implies that either
both photons 2 and 4 are horizontally polarized or both vertically
polarized. Therefore, by selecting those events where there is one
and only one photon in the output mode $4^{\prime}$, Alice and Bob
can obtain a conditioned four-photon Greenberger-Horne-Zeilinger
(GHZ) state \cite{pan01a}
\begin{equation}
\left\vert \Psi \right\rangle _{c}=\frac{1}{\sqrt{2}}\left( \left\vert
H_{1}\right\rangle \left\vert V_{2}^{\prime }\right\rangle \left\vert
V_{3}\right\rangle \left\vert V_{4}^{\prime }\right\rangle +\left\vert
V_{1}\right\rangle \left\vert H_{2}^{\prime }\right\rangle \left\vert
H_{3}\right\rangle \left\vert H_{4}^{\prime }\right\rangle \right) ,
\label{GHZ}
\end{equation}
with a probability of $2\left\vert \alpha \beta \right\vert
^{2}$. By further performing a polarization measurement on each
of the photons 3 and $4^{\prime}$ in the $+/-$ basis, where
$\left\vert\pm\right\rangle =1/\sqrt{2} \left( \vert
H\rangle\pm\vert V\rangle\right)$, Alice and Bob can then project
the photons 1 and $2^{\prime}$ onto the maximally entangled state
\cite{zhao01}
\begin{equation}
\left\vert \Psi^+ \right\rangle _{12^{\prime
}}=\frac{1}{\sqrt{2}}\left( \left\vert H_{1}\right\rangle
\left\vert V_{2^{\prime }}\right\rangle +\left\vert
V_{1}\right\rangle \left\vert H_{2^{\prime }}\right\rangle
\right).
\end{equation}

The scheme above deserves some further comments. Comparing with
the original concentration scheme by Bennett et al.
\cite{bennett96b}, the present scheme involves only two entangled
photon pairs. Hence, it is reachable using the techniques
developed in the recent four-photon experiment \cite{pan01a}.
Furthermore, the two pairs are required to have the same
coefficients at the same experimental trial, but for different
trials the coefficients can be different. In reality, this can
well be the case if the sources $S$ emit only two entangled pairs
at each time and thus only these two pairs experience the same
environment. Finally, but more importantly, the same idea can
also be used to implement a simplified version of quantum
repeaters.

To see how it works, suppose that there is an intermediate
station, Charlie, between Alice and Bob (see Fig. 1b). Further
suppose that Alice and Charlie share the pair (1,2) and, Charlie
and Bob share the pair (3,4). Here, the pairs (1,2) and (3,4) are
in the states as shown in equation (1). After the photon 4 passes
through the R$_{90}$, Charlie superposes the photons 2 and 4 at
the PBS. Similarly, by performing a polarization measurement on
each of the two outputs $2^{\prime}$ and $4^{\prime}$ in the
$+/-$ basis, Alice and Bob can thus obtain the maximally
entangled state
\begin{equation}
\left\vert \Psi^+ \right\rangle _{13}=\frac{1}{\sqrt{2}}\left(
\left\vert H_{1}\right\rangle \left\vert V_{3}\right\rangle
+\left\vert V_{1}\right\rangle \left\vert H_{3}\right\rangle
\right).
\end{equation}
In this protocol, both entanglement concentration and entanglement
swapping are accomplished in one step. Of course, if the two
entangled pairs have different coefficients, for example, the
pair (1,2) in the state $\alpha_{12} \left\vert H_{1}\right\rangle
\left\vert V_{2}\right\rangle +\beta_{12} \left\vert
V_{1}\right\rangle \left\vert H_{2}\right\rangle,$ and the pair
(3,4) in the state  $\alpha_{34} \left\vert H_{3}\right\rangle
\left\vert V_{4}\right\rangle +\beta_{34} \left\vert
V_{3}\right\rangle \left\vert H_{4}\right\rangle$, then the above
protocol is not applicable. In the latter case, if the
coefficients are known in advance, we can first apply local
filtering to each of the two pairs, then utilize entanglement
swapping to generate maximally entangled state $\left\vert \Psi^+
\right\rangle _{13}$. Thus, the task of a quantum repeater can
still be accomplished. It is worth noting that although the above
scheme is only a simplified variant, the scheme itself acquires
all the necessary features as a quantum repeater.

A schematic drawing of the experimental setup of entanglement
concentration is shown in Fig. 2a. In the experiment, we first
generate two photon pairs in the maximally entangled state
$\vert\Psi^{+}\rangle$ by type II parametric down-conversion
\cite{kwiat95}\ from an ultraviolet (UV) pulsed laser in a BBO
crystal. The UV pulse passing through the crystal creates the
first pair in modes 1 and 2. After retroflection, during its
second passage through the crystal, the laser pulse creates the
second pair in modes 3 and 4. The UV pulsed laser with a central
wavelength of 394nm has a pulse duration of 200fs, a repetition
rate of 76MHz, and an average power of 450mW. To prepare the
states in equation (1), we send photons 1 and 3 through the same
amount of Brewster's windows (BWs) with vertical axis tilted by
$56^{\circ}$. Due to the polarization-dependent reflectivity, the
transmitted photons are preferentially horizontally polarized.
Here, with one piece of BW, the transmission probability for
horizontal polarization is $T_{H}=0.98$, and for vertical one
$T_{V}=0.73$.

After photon 4 passes through the half wave plate R$_{90}$,
photons 2 and 4 are steered to the PBS where the path lengths of
the two photons have been adjusted such that they arrive
simultaneously. Through spectral filtering (F in Fig. 2) with a
$\Delta \lambda _{FWHM}=3.6$nm for all the four photons, the
coherence time of the photons was made to exceed the duration of
the UV pulse, thus making the two photons 2 and 4
indistinguishable in time \cite{marek95}. Furthermore,
fiber-coupled single-photon detectors have been used to ensure
good spatial mode overlap between 2 and 4. These arrangements
consequently lead to two-photon interference. Through the whole
experiment, the registered two-fold coincidence rate is about
$2.6\times 10^{4}$ per second before the Brewster's windows and
the PBS, which results in an overall four-fold coincidence of 8
per second, almost two orders of magnitude higher compared to the
recent four-photon experiment \cite{pan01a}.

In our experiment, we choose to insert one, two or four pieces of
BWs into each of the modes 1 and 3 to prepare three different
initial states (denoted by (a), (b), (c)). Furthermore, in order
to demonstrate the protocol works for complex coefficients
$\alpha$ and $\beta$ (i.e. for both amplitude and phase errors),
we also slightly tilt the birefrigent compensators (not shown in
the figure) in the modes 1 and 3 such that a $\pi/2$ relative
phase is introduced between the $\vert H\rangle\vert V\rangle$
and $\vert V\rangle\vert H\rangle$ components. Ideally, these
initial states should only contain $\vert H\rangle\vert V\rangle$
and $\vert V\rangle\vert H\rangle$ terms. However, in reality all
the four possible outcomes in the $H/V$ basis have been observed.
Correspondingly, the fractions of the four components are shown
in Figs. 3a, 3b and 3c. In all three caess, the signal-to-noise
ratio is better than 100:1. In addition, we also list in Table 1
the ratios between $\vert H\rangle\vert V\rangle$ and $\vert
V\rangle\vert H\rangle$.

If entanglement concentration works, then conditional on a $\vert
+\rangle\vert +\rangle$ coincidence detection in the modes 3 and
$4^{\prime}$, the remaining two photons 1 and $2^{\prime}$ would
be projected onto the maximally entangled state $\left\vert
\Psi^+ \right\rangle _{12^{\prime }}$. To verify this prediction,
we first measure the fractions of the photons 1 and $2^{\prime}$
in the $H/V$ basis after entanglement concentration. The
measurement results are shown in Figs. 3d, 3e and 3f,
respectively, from which we can see the signal-to-noise ratios
are larger than 20:1 for all three cases. Therefore, the
contributions of $\vert H\rangle\vert H\rangle$ and $\vert
V\rangle\vert V\rangle$ terms are negligible. At the same time,
in Table 1 we also list the ratios between $\vert H\rangle\vert
V\rangle$ and $\vert V\rangle\vert H\rangle$ after concentration.
Compared with the initial ratios, one can clearly see the
significant improvement in the relative intensity between $\vert
H\rangle\vert V\rangle$ and $\vert V\rangle\vert H\rangle$.

Showing that both $\vert H\rangle\vert V\rangle$ and $\vert
V\rangle\vert H\rangle$ appear with roughly the same probability
of $50\%$ is just a necessary but not sufficient condition to
verify the state $\left\vert \Psi^{+} \right\rangle _{12^{\prime
}}$, since the above observation is, in principle, both compliant
with $\left\vert \Psi \right\rangle _{12^{\prime }}$ and with a
statistical mixture of $\vert H\rangle\vert V\rangle$ and $\vert
V\rangle\vert H\rangle$. Thus, as a further test we have to
demonstrate that the two terms $\vert H\rangle\vert V\rangle$ and
$\vert V\rangle\vert H\rangle$ are indeed in a coherent
superposition. To do so, we further perform a polarization
analysis in the $+/-$ basis. Transforming $\left\vert \Psi^+
\right\rangle _{12^{\prime }}$ to the $+/-$ basis yields an
expression containing only $\vert +\rangle\vert +\rangle$ and
$\vert -\rangle\vert -\rangle$, but no other terms like $\vert
+\rangle\vert -\rangle$ and $\vert -\rangle\vert +\rangle$. As a
test for coherence we can now check the presence or absence of
various components. In Fig. 4 we compare the $\vert +\rangle\vert
+\rangle$ and $\vert -\rangle\vert +\rangle$ count rates as a
function of the pump delay mirror position conditional on
detecting a $\vert +\rangle\vert +\rangle$ coincidence in the
modes 3 and $4^{\prime}$. At zero delay -- photons $2$ and $4$
arrive at the PBS simultaneously -- the latter component is
suppressed with a visibility of $0.83\pm 0.04$, $0.80\pm 0.05$
and $0.80\pm 0.04$, respectively.

The high visibility observed in the $+/-$ basis implies a
violation of a suitable Bell's inequality for the photons 1 and
$2^{\prime}$. According to the inequality of
Clauser-Horne-Shimony-Holt (CHSH) \cite{clauser69}, there is a
constraint on the value of $S$, a combination of four polarization
correlation probabilities -- two possible analysis settings for
each photon. If $|S|\leq 2$, no quantum entanglement is necessary
to explain the correlations, that is, some local realistic model
can reproduce them. For the maximally entangled state $\left\vert
\Psi^{+} \right\rangle _{12^{\prime }}$, the maximum value of $S$
is $2\sqrt{2}$. Therefore, the value of $S$ is a figure of merit
for the quality of the entanglement between the photons 1 and
$2^{\prime}$. In the experiment, we also measured the $S$ value
for all three cases. The integration time is about $10^{3}$
seconds for each possible outcomes in the first two cases, and
$2\times10^{3}$ seconds in the third case. The CHSH inequality was
violated by more than 5 standard deviations in all cases; see
Table 1. Since the entanglement visibility $V$ satisfies the
relation $V=S/2\sqrt{2}$, we can calculate the visibility $V$
from the observed value $S$ and further estimate the fidelity for
the photons 1 and $2^{\prime}$ to be in the state $\left\vert
\Psi^{+} \right\rangle _{12^{\prime }}$. As shown in Table 1, the
fidelity obtained is larger than $0.93$ in all cases.

In order to show our experiment also provides an realization of a
quantum repeater, we reassemble the setup as shown in Fig. 2b. We
first consider the case where the two photon pairs (1,2) and
(3,4) are in the same initial state. In this case, as discussed
in equation (5), conditional on detecting a $\vert +\rangle \vert
+\rangle$ coincidence in $2^{\prime}$ and $4^{\prime}$, the
photons 1 and 3 shared by Alice and Bob will be left in the state
$\vert\Psi^{+}\rangle_{13}$. In the first repeater experiment, we
choose the initial state (b) to verify this prediction. This is
done by measuring the fractions of the photons 1 and 3 in both
$H/V$ and $+/-$ bases. While HV and VH components are observed
with roughly the same probability of $50\%$, both HH and VV terms
are negligible. In addition, an interference visibility of
$0.81\pm0.04$ is observed in the $+/-$ basis. The $S$ value for
the photons 1 and 3 is measured to be $2.44\pm 0.08$, from which
the fidelity for the photons 1 and 3 to be in the state
$\left\vert \Psi^{+} \right\rangle _{13}$ is about $0.93\pm 0.04$.

We now consider the case that the two photon pairs (1,2) and
(3,4) are in different initial states. In the second repeater
experiment, we first prepare the pair (1,2) in the state (a) and
the pair (3,4) in the state (b). We then perform local filtering
to convert the pairs (1,2) and (3,4) into the same maximally
entangled state $\vert\Psi^{+}\rangle$. Finally, we utilize
entanglement swapping to generate the desired state $\left\vert
\Psi^{+} \right\rangle _{13}$ between Alice and Bob. Again, this
is achieved by detecting a $\vert +\rangle \vert +\rangle$
coincidence in $2^{\prime}$ and $4^{\prime}$. To verify the
success of our quantum repeater protocol, we also measured the $S$
value for the photons 1 and 3, which is $2.52\pm 0.10$. This
implies that the fidelity for the photons 1 and 3 to be in the
state $\left\vert \Psi^{+} \right\rangle _{13}$ is about $0.95\pm
0.05$.

It should be mentioned that in the above repeater experiments both
entanglement purification and entanglement swapping play a
crucial role in generating maximally entangled states between
Alice and Bob. Without applying entanglement concentration or
local filtering, the operation of entanglement swapping alone will
only generate a non-maximally entangled state, say,
$\alpha^{2}\vert H\rangle\vert V\rangle+\beta^{2}\vert
H\rangle\vert V\rangle$ in the first repeater experiment. Such a
state acquires an entanglement fidelity even worse than the
original pairs. The necessity of both entanglement concentration
(or local filtering) and entanglement swapping implies that both
experiments present a faithful demonstration of a quantum
repeater. Furthermore, after taking into account the imperfection
of the $\vert\Psi^{+}\rangle$ state preparation in the scond
experiment, we can estimate the PBS accuracy to interfere two
independent photons from the experimental visibility for
entanglement swapping, which is about $98\%$. This clearly
surpasses the strict precision requirement of local operations for
the multi-stage realization of quantum repeaters \cite{briegel98}.
Therefore, the techniques developed for our entanglement
concentration and quantum repeater experiment provide a practical
toolbox for future realization of long-distance quantum
communication.

\section*{Acknowledgements}

This work was supported by the National Natural Science
Foundation of China, the Chinese Academy of Sciences and the
National Fundamental Research Program (under Grant No.
2001CB309303).

* To whom correspondence should be addressed.
E-mail:pan@ap.univie.ac.at

\newpage

\begin{table}
\begin{tabular}{ccccc}
\multicolumn{5}{l}{\bf Table 1 Summary of entanglement concentration data for different BWs}\\
\hline Number of BWs & Pre-concentration & After concentration & $S$ values & Fidelities \\
\hline
1 & 1.41:1 & 1.08:1 & 2.58$\pm0.07$ & 0.96$\pm0.04$\\
2 & 1.72:1 & 1.09:1 & 2.43$\pm0.08$ & 0.93$\pm0.04$\\
4 & 3.15:1 & 1.10:1 & 2.42$\pm0.08$ & 0.93$\pm0.04$\\
\hline
\end{tabular}
\end{table}

\noindent \textbf{Figure Captions:}

Figure captions:

Figure $1$: Schematic drawings showing the principles of
entanglement concentration ({\bf a}) and quantum repeater ({\bf
b}). {\bf a}, The sources S emit two entangled photon pairs into
the modes (1,2) and (3,4). The two pairs are originally in the
same unknown non-maximally entangled state. One member of each
pair is then sent to Alice, and the other one to Bob. After
photon 4 passes through the half wave plate R$_{90}$, Bob
superposes the photons at the polarizing beam splitter PBS. By
performing a polarization measurement on each of the photons in
mode 3 and $4^{\prime}$ in the $+/-$ basis, Alice and Bob can
project the photons 1 and $2^{\prime}$ onto a maximally entangled
state. {\bf b}, Alice and Charlie share the pair (1,2) and, Bob
and Charlie share the pair (3,4). Interfering the photons 2 and 4
at the PBS and performing a subsequent polarization measurement
on each of the modes $2^{\prime}$ and $4^{\prime}$ in the $+/-$
basis, Charlie can thus generate a maximally entangled state
between Alice and Bob (see text).

Figure $2$: The experimental setups for entanglement
concentration ({\bf a}) and quantum repeater ({\bf b}). {\bf a},
A UV pulse passes through a BBO twice to generate two entangled
photon pairs. By inserting one, two or four pieces of Brewster's
windows (BWs) into each of the modes 1 and 3, we can prepare three
different nonmaximally entangled states. After the photons 2 and
4 pass through the PBS, conditional on a two-fold coincidence
detection behind the two $45^{\circ}$ polarizers Pol in the modes
3 and $4^{\prime}$, the photons $1$ and $2^{\prime }$ will be in a
maximally entangled state. {\bf b}, If the pairs (1,2) and (3,4)
are in the same nonmaximally entangled state, then detecting a
two-fold coincidence behind the two $45^{\circ}$ polarizers Pol
in the modes $2^{\prime}$ and $4^{\prime}$ is sufficient to
project the photons 1 and 3 onto the desired maximally entangled
state. Otherwise, additional half wave plates and BWs are needed
in the modes 1 and 3 to generate the desired entangled state
between Alice and Bob.

Figure $3$: Experimental results demonstrating the entanglement
concentration. {\bf a}, {\bf b}, and {\bf c} show the fractions
of the nonmaximally entangled state in the $H/V$ basis, when
inserting one, two and four BWs into the modes 1 and 3. {\bf d},
{\bf e}, and {\bf f} show the corresponding fractions of the
photons 1 and $2^{\prime}$ in the $H/V$ basis after entanglement
concentration. The data clearly confirm that our concentration
protocol significantly improved the relative intensity between
$\vert H\rangle\vert V\rangle$ and $\vert H\rangle\vert V\rangle$
components.

Figure $4$: Experimental results confirming the entanglement
between the photons 1 and $2^{\prime}$. The two-fold coincidence
count rate $\vert +\rangle\vert +\rangle$ and $\vert -\rangle\vert
+\rangle$  is shown as a function of the pump delay mirror
position under the condition that a $\vert +\rangle\vert +\rangle$
coincidence is registered in the modes $3$ and $4^{\prime}$.

\newpage

\begin{figure}[ht]
\begin{center}
\includegraphics[width=0.8\columnwidth]{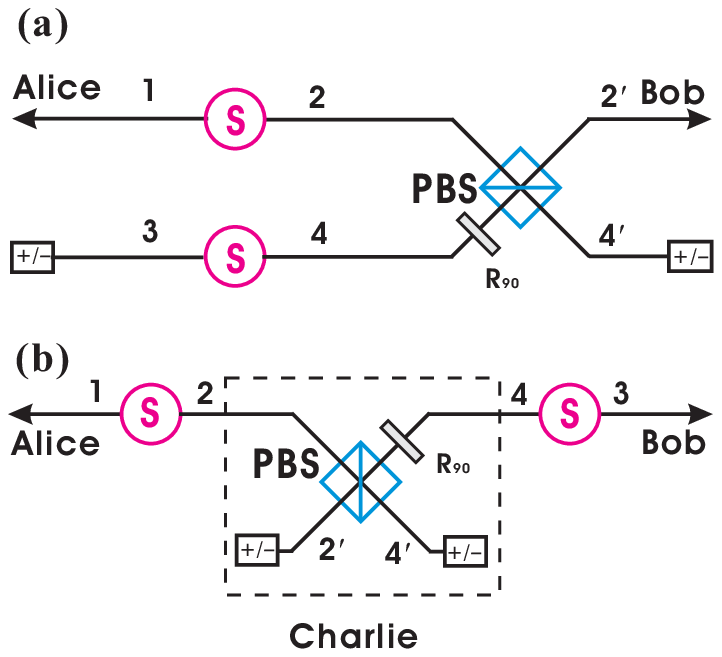}
\caption{}
\end{center}
\end{figure}

\newpage

\begin{figure}[ht]
\begin{center}
\includegraphics[width=0.8\columnwidth]{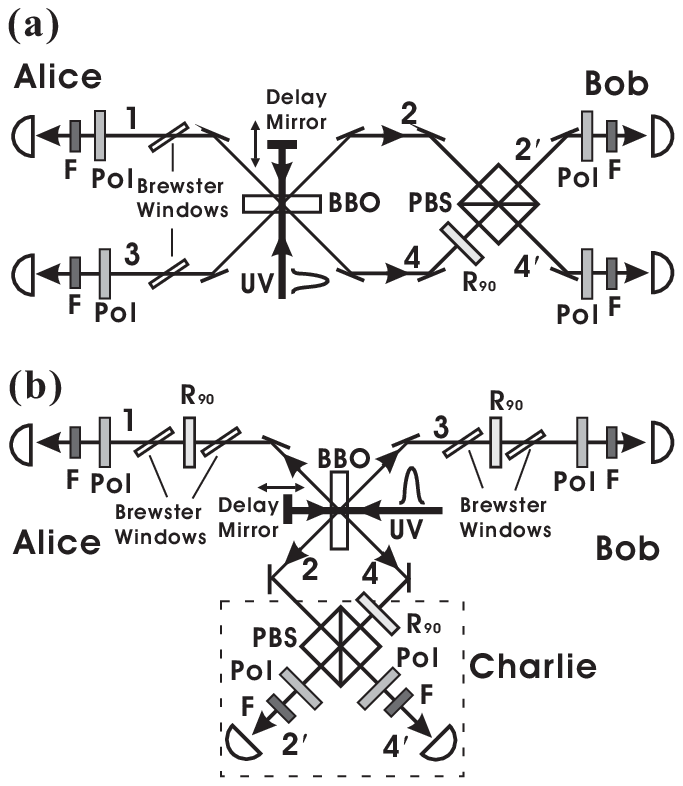}
\caption{}
\end{center}
\end{figure}

\newpage

\begin{figure}[ht]
\begin{center}
\includegraphics[width=0.7\columnwidth]{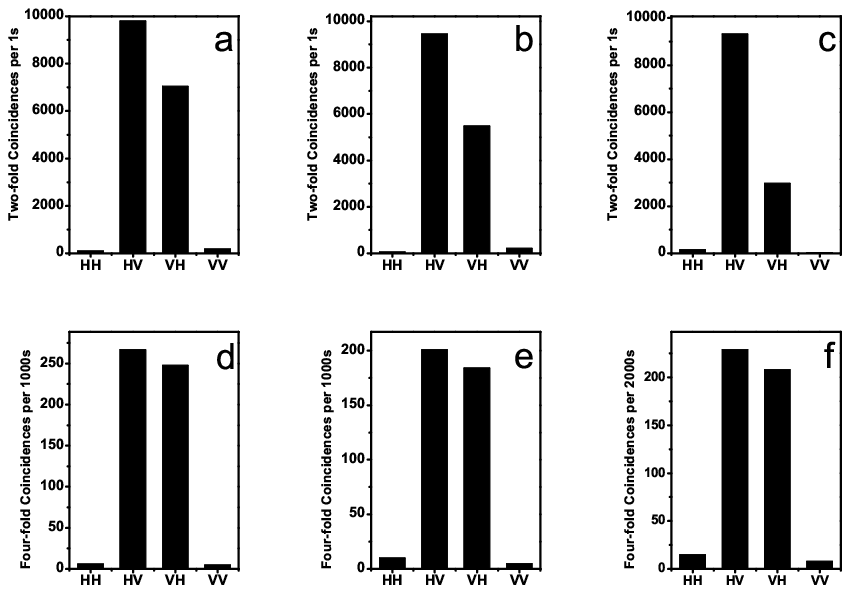}
\caption{}
\end{center}
\end{figure}

\newpage

\begin{figure}[ht]
\begin{center}
\includegraphics[width=0.8\columnwidth,clip=true]{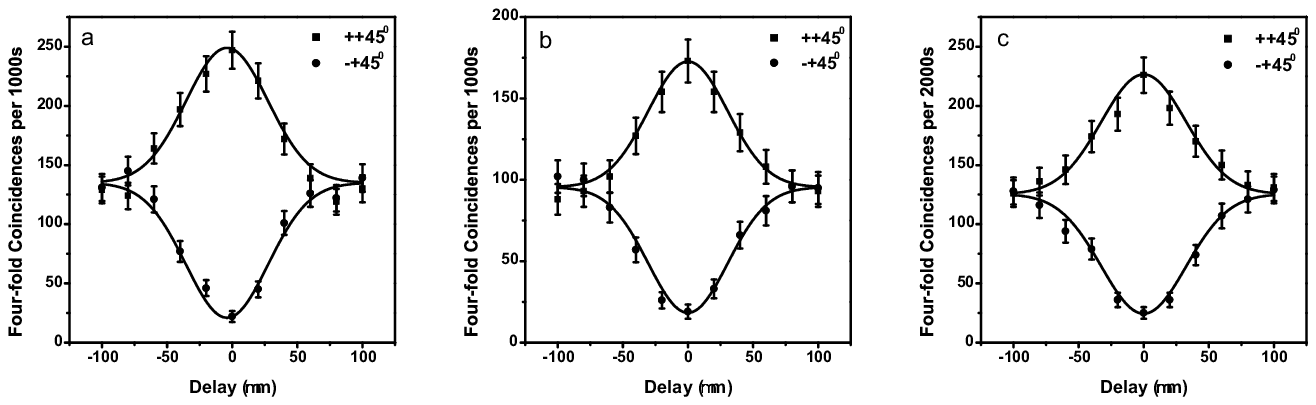}
\caption{}
\end{center}
\end{figure}

\end{document}